\documentclass[11pt, reprint,showkeys, amsmath,amssymb, aps, nofootinbib, floatfix]{revtex4-2}
\DeclareUnicodeCharacter{2212}{\textendash}
\usepackage{graphicx,subfigure}
\graphicspath{{figures/}}
\usepackage[export]{adjustbox}
\usepackage{dcolumn}
\usepackage{tabularx}
\usepackage{bm}
\usepackage{xcolor}
\usepackage{braket}
\usepackage{slashed,cancel}
\usepackage{multirow}
\usepackage{float}
\usepackage{capt-of}
\usepackage{comment}
\usepackage{mathrsfs}
\usepackage{enumitem}
\usepackage{booktabs}
\usepackage[colorlinks=true,linkcolor=blue,urlcolor=blue,citecolor=blue]{hyperref}

\allowdisplaybreaks

\begin{document}
\preprint{APS/123-QED}

\title{Multiparameter Quantum Estimation and Degeneracy Structure in Three-Flavor Neutrino Oscillations}

\author{Bhavna Yadav}
\email{bhavnayadav04@ustc.edu.cn} %

\author{Amir Subba }
\email{ amirsubba@ustc.edu.cn} %
	
\author{Yu Shi }
\email{yu\_shi@ustc.edu.cn} %
\affiliation{Wilczek Quantum Center, Shanghai Institute for Advanced Studies USTC, Shanghai 201315, China}
\affiliation{University of Science and Technology of China, Hefei 230026, China }

\begin{abstract}
Achieving precision measurements of neutrino oscillation parameters and understanding parameter degeneracies remain central challenges in neutrino physics. We investigate three-flavor neutrino oscillations within the framework of quantum estimation theory using the quantum Fisher information matrix (QFIM). We analyze all six independent elements of the QFIM for the parameters $\theta_{23}$, $\delta_{\mathrm{CP}}$, and $\Delta m^2_{31}$, together with the corresponding quantum Cram\'er--Rao bounds and normalized parameter correlations. To connect the intrinsic quantum sensitivity of the neutrino state with experimentally accessible information, we further compare the QFI with the classical Fisher information (CFI) associated with flavor-probability measurements. We find that the difference between the quantum and classical information bounds is strongly dependent on the oscillation parameter and $L/E$, demonstrating that the efficiency with which the available quantum information is accessed depends on the measurement. We also investigate probability-degenerate solutions in the $(\theta_{23},\delta_{\mathrm{CP}})$ parameter space. Although different parameter sets can produce identical $\nu_\mu\rightarrow\nu_e$ appearance probabilities at specific $L/E$ values, their corresponding quantum states can remain distinct, as demonstrated by the quantum fidelity. The QFIM differences further reveal distinct local quantum-information geometries for these probability-degenerate solutions, while the corresponding CFIM differences show a parameter and $L/E$ dependent degree of distinguishability accessible through the flavor-probability measurement. Our results demonstrate that QFI and CFI provide complementary information for neutrino parameter estimation, with the QFIM serving as a measurement-independent quantum benchmark and the CFI characterizing the information accessible through a specified measurement.
\end{abstract}

\maketitle

\section{Introduction}
\label{sec:intro}

One of the most profound and enduring puzzles within the Standard Model (SM)~\cite{Weinberg:1967tq} of particle physics concerns the origin of neutrino mass. The SM, in its minimal formulation, predicts exactly massless neutrinos, which is a consequence of the absence of right-handed neutrino fields and, correspondingly, the lack of any gauge-invariant Yukawa coupling that could otherwise generate a mass term upon electroweak symmetry breaking. This theoretical prediction, however, stands in sharp tension with experimental observations.

The observed deficit in the detected flux of electron neutrinos originating from the solar core~\cite{Davis:1968cp}, known as the \emph{solar neutrino problem}, provided the first compelling experimental evidence incompatible with the SM prediction of massless neutrinos. This deficit finds a natural resolution within the framework of neutrino oscillations, wherein the flavor eigenstates $\ket{\nu_\alpha}$ ($\alpha=e,\mu,\tau$) are expressed as coherent superpositions of mass eigenstates $\ket{\nu_i}$ ($i=1,2,3$) through the unitary Pontecorvo--Maki--Nakagawa--Sakata (PMNS)~\cite{Maki:1962mu,Pontecorvo:1967fh,Gribov:1968kq} mixing matrix $U_{\alpha i}$, i.e., $\ket{\nu_\alpha}=\sum_i U_{\alpha i}^*\ket{\nu_i}$. As a consequence, a neutrino produced in a definite flavor state undergoes coherent flavor evolution during propagation, a dynamical effect that is contingent upon non-vanishing mass-squared splittings, $\Delta m^2_{ij}\equiv m_i^2-m_j^2\neq0$. Experimental evidence from atmospheric, solar, reactor, and accelerator neutrino sources~\cite{Super-Kamiokande:1998kpq,SNO:2002tuh,KamLAND:2004mhv,Super-Kamiokande:2016yck} has since firmly established the existence of neutrino oscillations, unambiguously implying that neutrinos have non-zero mass.

The oscillation framework is characterized by three mixing angles, two independent mass-squared differences, and the leptonic CP-violating phase~\cite{Maki:1962mu,Bilenky:1978nj}. Precision measurements from solar, atmospheric, reactor, and accelerator experiments have now constrained the three-flavor oscillation parameters with remarkable accuracy~\cite{Capozzi:2021fjo,Esteban:2024eli}. Despite this progress, several important challenges remain, particularly the precise determination of the leptonic CP phase, the resolution of the $\theta_{23}$ octant ambiguity, and the identification of the neutrino mass ordering~\cite{Esteban:2024eli,Agarwalla:2016fkh,Rahaman:2022rfp,DiLodovico:2023jgr,T2K:2025wet}. In particular, different combinations of oscillation parameters can lead to identical or nearly identical transition probabilities, giving rise to well-known degeneracies involving the atmospheric mixing angle, the CP phase, and the neutrino mass ordering~\cite{Barger:2001yr,Minakata:2001qm,Kajita:2006bt,Hiraide:2006vh,Minakata:2010zn,Ghosh:2015ena,Nath:2015kjg,Das_2015,Agarwalla:2016fkh,Giganti:2017fhf,Sugama:2023sjf}. Such degeneracies represent a major obstacle in the interpretation of oscillation data and in the optimization of future neutrino experiments.

The resolution of parameter degeneracies has therefore become one of the central objectives of current and future neutrino oscillation experiments. Considerable effort has been devoted to mitigating these degeneracies through the use of multiple baselines, complementary oscillation channels, combined analyses of neutrino and antineutrino data, and synergistic studies involving different experiments~\cite{Agarwalla:2016fkh,Rahaman:2022rfp,DiLodovico:2023jgr}. While these approaches can significantly improve parameter determination, degeneracies continue to limit the precision with which the oscillation parameters can be extracted and remain a major challenge for future precision measurements. Consequently, it is important to investigate whether information beyond the transition probabilities can provide additional insight into parameter sensitivity and the structure of these degeneracies. Within the conventional statistical framework, the sensitivity to oscillation parameters is quantified by the classical Fisher information (CFI), which characterizes the information extracted from measurement outcomes. However, the CFI depends explicitly on the chosen measurement strategy and therefore does not characterize the ultimate information content of the underlying quantum state.

Quantum Fisher information (QFI) provides a natural extension of the classical Fisher information by quantifying the intrinsic sensitivity of a quantum state to variations of physical parameters. Unlike its classical counterpart, the QFI is independent of a specific measurement scheme and determines the ultimate precision bound through the quantum Cram\'er--Rao bound~\cite{Fisher:1925mvw,Cramer1946,Jaynes:1957zza,Helstrom:1969fri,Paris:2008zgg,Lu:2010ktl,Toth:2014msl,Pezze:2014gld,Liu:2019xfr}. In the context of neutrino oscillations, QFI has previously been employed to study the precision with which the CP phase and other oscillation parameters can be estimated~\cite{Nogueira:2016qsk,Ignoti:2025rxr,Yadav:2026lsx,Frugiuele:2026yeq,Farooq:2026eap,Chundawat:2026jjd,Chundawat:2026lcm}. Nevertheless, existing studies have predominantly focused on individual-parameter sensitivity, while a systematic multiparameter analysis based on the full QFIM remains less explored within the three-flavor oscillation framework. In particular, the role of off-diagonal QFIM elements, normalized parameter correlations, and their impact on simultaneous parameter estimation has not yet been fully investigated. Moreover, the relation between the measurement-independent QFI and the experimentally accessible CFI in neutrino oscillations provides an important connection that warrants further investigation.

In view of these considerations, in this work we develop a multiparameter quantum-estimation framework for three-flavor neutrino oscillations based on the full QFIM. Focusing on the parameter set $(\theta_{23},\delta_{\mathrm{CP}},\Delta m^2_{31})$, we investigate the individual parameter sensitivities, the correlations among parameters, and the resulting multiparameter precision. We further compare the QFI with the CFI associated with flavor-probability measurements to quantify how the experimentally accessible information approaches the quantum-information limit. This comparison allows us to distinguish the intrinsic information content of the neutrino quantum state from the information accessible through a specific measurement.

We also revisit parameter degeneracies from a quantum-information perspective. While degeneracies are conventionally identified through identical or nearly identical oscillation probabilities, the corresponding quantum states need not possess identical local information geometry. We therefore consider representative parameter sets that produce degenerate $\nu_\mu\rightarrow\nu_e$ appearance probabilities and examine their quantum fidelity, QFIM differences, and corresponding CFIM differences. Our analysis shows that probability-degenerate solutions can possess distinct local quantum-information geometries, while the corresponding differences in the accessible classical information depend on the parameter direction and the chosen measurement. Thus, the equality of an appearance probability does not necessarily imply equality of the underlying quantum-state information structure or of the local classical information accessible through the measurement.

The paper is organized as follows. In Sec.~\ref{sec:formalism}, we briefly review the formalism of three-flavor neutrino oscillations and the quantum Fisher information matrix. In Sec.~\ref{sec:individual}, we discuss the diagonal QFI elements and the corresponding quantum Cram\'er--Rao bounds, together with their comparison to the classical Fisher information. In Sec.~\ref{sec:multi}, we investigate the off-diagonal QFIM and CFIM elements, the corresponding normalized parameter correlations, and the impact of multiparameter correlations on the attainable precision. In Sec.~\ref{sec:octant}, we study probability-degenerate solutions using quantum fidelity and compare the local quantum and classical information structures through the QFIM and CFIM differences. Finally, in Sec.~\ref{sec:conclusion}, we summarize our main findings and discuss possible directions for future work.

{\color{black}\section{Formalism: Neutrino Evolution and Quantum Fisher Information Matrix}
\label{sec:formalism}

A neutrino produced in a flavor state $|\nu_\alpha\rangle$ can be expressed as a coherent superposition of the mass eigenstates $|\nu_i\rangle$ as
\begin{align}
|\nu_\alpha\rangle
=
\sum_i U_{\alpha i}^{*}|\nu_i\rangle,
\end{align}
where $U$ denotes the Pontecorvo--Maki--Nakagawa--Sakata (PMNS) mixing matrix. In the standard parametrization, the PMNS matrix can be written as~\cite{Valle:2006vb}
\begin{align}
U &=&
\begin{pmatrix}
1 & 0 & 0\\
0 & c_{23} & s_{23}\\
0 & -s_{23} & c_{23}
\end{pmatrix}
\begin{pmatrix}
c_{13} & 0 & s_{13}e^{-i\delta_{\rm CP}}\\
0 & 1 & 0\\
-s_{13}e^{i\delta_{\rm CP}} & 0 & c_{13}
\end{pmatrix} \nonumber 
\\
&&\times 
\begin{pmatrix}
c_{12} & s_{12} & 0\\
-s_{12} & c_{12} & 0\\
0 & 0 & 1
\end{pmatrix},
\end{align}
where $c_{ij}\equiv\cos\theta_{ij}$ and $s_{ij}\equiv\sin\theta_{ij}$. The matrix is characterized by three mixing angles $(\theta_{12},\theta_{13},\theta_{23})$ and the Dirac CP-violating phase $\delta_{\rm CP}$.

During propagation, the neutrino state evolves according to the effective Hamiltonian. In the mass basis, and in the relativistic limit, the vacuum Hamiltonian is given by
\begin{align}
H_{\rm vac}
=
\frac{1}{2E}
\begin{pmatrix}
0 & 0 & 0\\
0 & \Delta m^2_{21} & 0\\
0 & 0 & \Delta m^2_{31}
\end{pmatrix},
\end{align}
where an overall term proportional to the identity has been omitted since it contributes only an unobservable global phase. In the flavor basis, the corresponding Hamiltonian is
\begin{align}
H_f = U H_{\rm vac} U^\dagger .
\end{align}
For the vacuum analysis considered in this work, the evolution operator is therefore
\begin{align}
S(L)=e^{-iH_f L},
\end{align}
and the evolved neutrino state is
\begin{align}
|\nu_\alpha(L)\rangle
=
S(L)|\nu_\alpha\rangle .
\end{align}
Equivalently, in the mass basis, each mass eigenstate acquires a phase during propagation \cite{Ohlsson:1999xb,Giunti:2007ry},
\begin{align}
|\nu_\alpha(L)\rangle
=
\sum_i U_{\alpha i}^{*}e^{-iE_iL}|\nu_i\rangle .
\end{align}

The probability of detecting the neutrino in a flavor state $|\nu_\beta\rangle$ is given by
\begin{align}
P_{\alpha\beta}(L)
=
\left|\langle\nu_\beta|\nu_\alpha(L)\rangle\right|^2 .
\end{align}
In conventional neutrino-oscillation analyses, the oscillation parameters are inferred from these transition probabilities. However, different combinations of oscillation parameters can lead to identical or nearly identical probabilities, giving rise to parameter degeneracies.

To investigate parameter sensitivity at the level of the underlying quantum state, we employ the quantum Fisher information matrix (QFIM). For a pure quantum state $|\psi(\boldsymbol{\theta})\rangle$ that depends on a set of parameters $\boldsymbol{\theta}=(\theta_1,\theta_2,\ldots)$, the QFIM is defined as
\begin{align}
\mathcal{F}_{ij}
=
4\,\mathrm{Re}\!\left[
\langle\partial_i\psi|\partial_j\psi\rangle
-
\langle\partial_i\psi|\psi\rangle
\langle\psi|\partial_j\psi\rangle
\right],
\end{align}
where
\begin{align}
\partial_i\equiv\frac{\partial}{\partial\theta_i}.
\end{align}
In this work, we consider the parameter vector
\begin{equation}
\boldsymbol{\Theta}
=
(\theta_{23},\,\delta_{\rm CP},\,\Delta m^2_{31}),
\end{equation}
which leads to a $3\times3$ QFIM describing both the individual parameter sensitivities and the correlations among the parameters. Explicitly,
\begin{equation}
\mathcal{F}(\boldsymbol{\Theta})=
\begin{pmatrix}
\mathcal{F}_{\theta_{23}\theta_{23}} &
\mathcal{F}_{\theta_{23}\delta_{\rm CP}} &
\mathcal{F}_{\theta_{23}\Delta m^2_{31}}
\\[2mm]
\mathcal{F}_{\delta_{\rm CP}\theta_{23}} &
\mathcal{F}_{\delta_{\rm CP}\delta_{\rm CP}} &
\mathcal{F}_{\delta_{\rm CP}\Delta m^2_{31}}
\\[2mm]
\mathcal{F}_{\Delta m^2_{31}\theta_{23}} &
\mathcal{F}_{\Delta m^2_{31}\delta_{\rm CP}} &
\mathcal{F}_{\Delta m^2_{31}\Delta m^2_{31}}
\end{pmatrix}.
\end{equation}

The diagonal elements $\mathcal{F}_{ii}$ quantify the intrinsic quantum sensitivity to individual parameters, while the off-diagonal elements $\mathcal{F}_{ij}$ characterize the correlations between different parameters in the multiparameter estimation problem. The corresponding quantum Cram\'er--Rao bound (QCRB) provides a lower bound on the covariance matrix of unbiased estimators,
\begin{align}
\mathrm{Cov}(\boldsymbol{\Theta})
\succeq
\mathcal{F}^{-1},
\end{align}
so that the marginal standard deviation of the parameter $\theta_i$ satisfies
\begin{align}
\sigma(\theta_i)
\geq
\sqrt{(\mathcal{F}^{-1})_{ii}} .
\end{align}
Thus, in the multiparameter case, the full inverse QFIM must be considered when evaluating the marginal precision, rather than using only the inverse of the corresponding diagonal element.

Since the QFI represents the maximum information that can, in principle, be extracted from the quantum state, it is useful to compare it with the classical Fisher information associated with a specific measurement. For a set of measurement outcomes labeled by $\beta$ with probabilities $P_\beta(\boldsymbol{\Theta})$, the classical Fisher information matrix (CFIM) is defined as
\begin{align}
\mathcal{F}^{C}_{ij}
=
\sum_{\beta}
\frac{1}{P_\beta}
\frac{\partial P_\beta}{\partial\theta_i}
\frac{\partial P_\beta}{\partial\theta_j},
\end{align}
where the sum runs over the possible flavor-detection outcomes. In the present analysis, we consider the flavor-probability measurement with
\begin{align}
\beta=e,\mu,\tau,
\end{align}
so that
\begin{align}
\mathcal{F}^{C}_{ij}
=
\sum_{\beta=e,\mu,\tau}
\frac{1}{P_{\alpha\beta}}
\frac{\partial P_{\alpha\beta}}{\partial\theta_i}
\frac{\partial P_{\alpha\beta}}{\partial\theta_j}.
\end{align}
The CFIM depends explicitly on the chosen measurement, whereas the QFIM characterizes the intrinsic information content of the quantum state. Consequently, the QFIM provides a measurement-independent quantum benchmark, while the CFIM quantifies the information accessible through the specified flavor-probability measurement.

\begin{table}[!htb]
\centering
\renewcommand{\arraystretch}{1.4}
\begin{tabular}{|c|c|c|}
\hline
\textbf{Parameter} &
\multicolumn{1}{c|}{\textbf{Normal Ordering}} &
\multicolumn{1}{c|}{\textbf{Inverted Ordering}} \\

& \textbf{Best fit $\pm 1\sigma$} &
\textbf{Best fit $\pm 1\sigma$} \\

\hline
$\theta_{12}\,[^\circ]$
& $33.68^{+0.73}_{-0.70}$
& $33.68^{+0.73}_{-0.70}$ \\

$\theta_{23}\,[^\circ]$
& $43.3^{+1.0}_{-0.8}$
& $47.9^{+0.7}_{-0.9}$ \\

$\theta_{13}\,[^\circ]$
& $8.56^{+0.11}_{-0.11}$
& $8.59^{+0.11}_{-0.11}$ \\

$\delta_{\rm CP}\,[^\circ]$
& $212^{+26}_{-41}$
& $274^{+25}_{-23}$ \\

$\Delta m^2_{21}\,[10^{-5}\,\mathrm{eV}^2]$
& $7.49^{+0.19}_{-0.19}$
& $7.49^{+0.19}_{-0.19}$ \\

$\Delta m^2_{31}\,[10^{-3}\,\mathrm{eV}^2]$
& $+2.513^{+0.021}_{-0.019}$
& $-2.484^{+0.020}_{-0.020}$ \\

\hline
\end{tabular}
\caption{
Best-fit values and $1\sigma$ ranges of the three-flavor neutrino
oscillation parameters for IC24 with SK-atm data from the NuFit-6.0
global analysis~\cite{Esteban:2024eli}.
}
\label{tab:nufit_bestfit_1sigma}
\end{table}}

\begin{figure*}[!htb]
    \centering
    \includegraphics[width=0.49\textwidth]{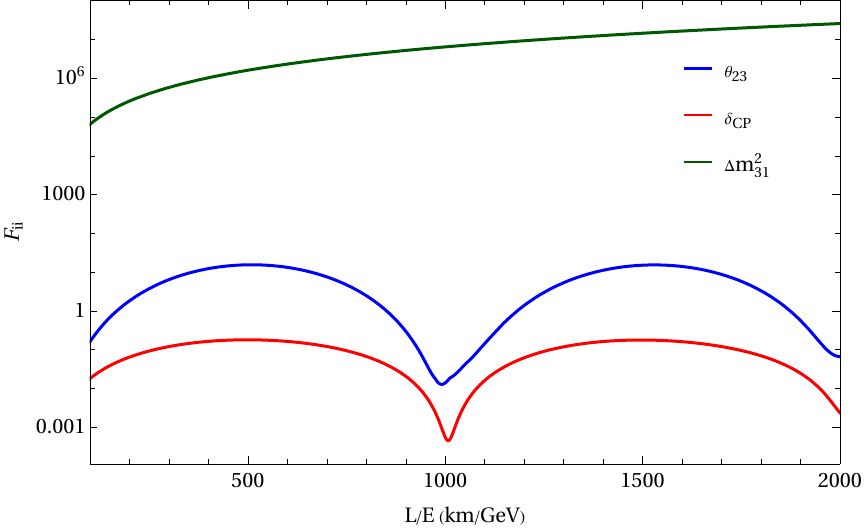}
    \includegraphics[width=0.49\textwidth]{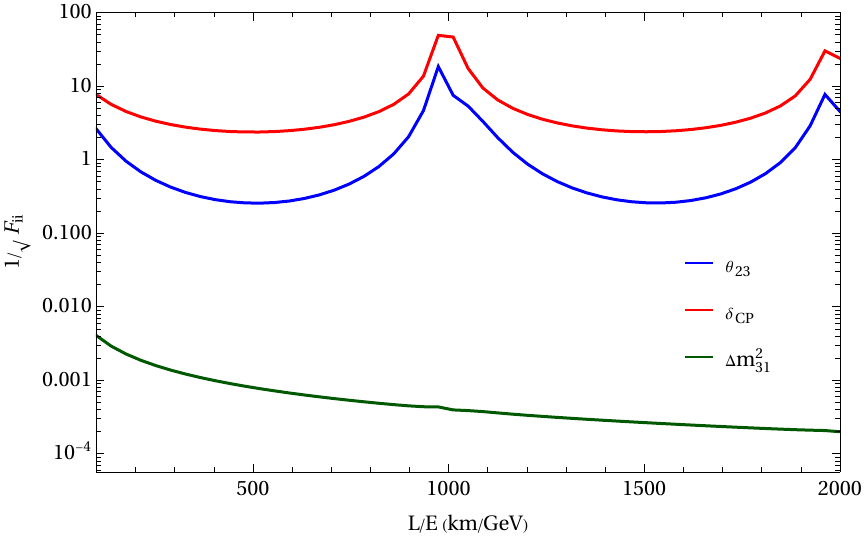}
   \caption{Quantum Fisher information of the diagonal elements of the QFIM (left panel) and the corresponding quantum Cram\'er-Rao bounds (right panel) for the oscillation parameters $\theta_{23}$, $\delta_{\mathrm{CP}}$, and $\Delta m_{31}^{2}$ as function of baseline length-to-energy ratio. 
}
    \label{fig:1}
\end{figure*}

\begin{figure*}[!htb]
    \centering
    \includegraphics[width=0.32\linewidth,height=4cm]{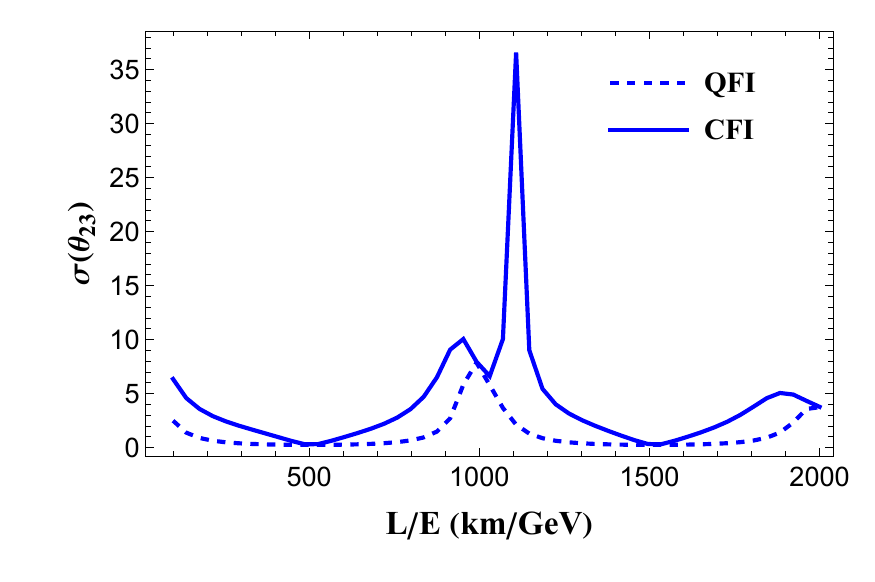}
    \includegraphics[width=0.32\linewidth,height=4cm]{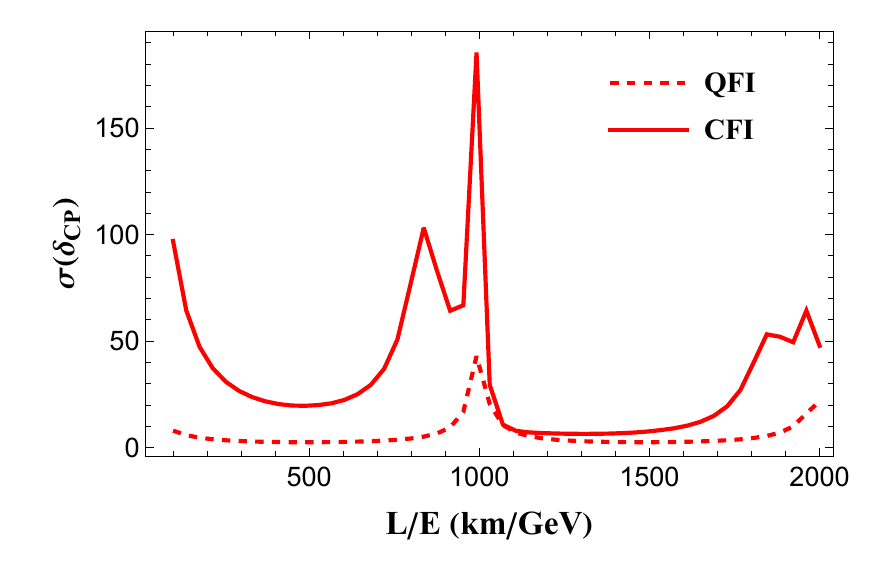}
    \includegraphics[width=0.32\linewidth,height=4cm]{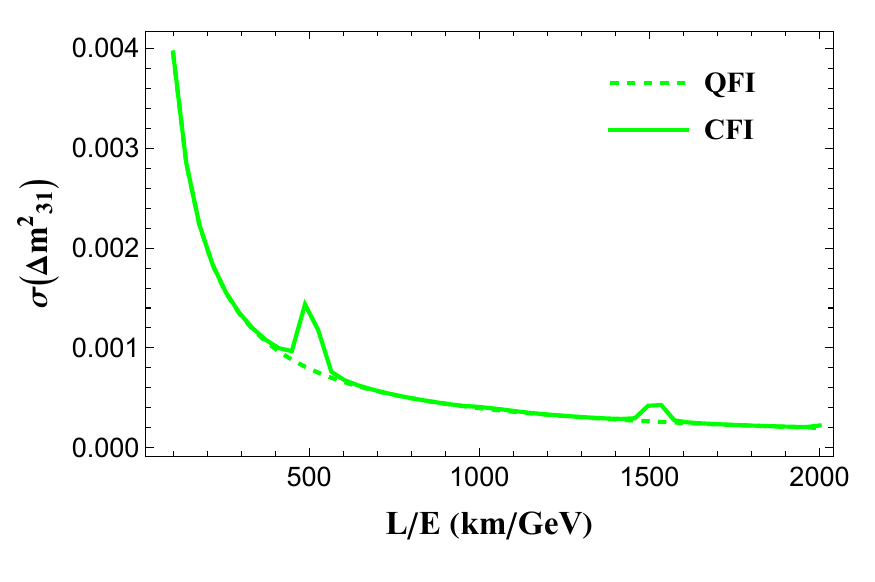}
   \caption{
Comparison of the standard deviations obtained from the diagonal elements of the quantum Fisher information matrix (QFI, dashed curves) and the classical Fisher information matrix (CFI, solid curves) for the oscillation parameters $\theta_{23}$, $\delta_{\mathrm{CP}}$, and $\Delta m^2_{31}$ as functions of $L/E$.
}
    \label{fig:2}
\end{figure*}


{\color{black}\section{Individual Parameter Sensitivity and Quantum-Limited Precision}
\label{sec:individual}

Fig.~\ref{fig:1} shows the QFI of the diagonal elements of the QFIM (left panel) and the corresponding quantum Cram\'er--Rao bounds (QCRBs) (right panel) as functions of the baseline-to-energy ratio $(L/E)$. Throughout the analysis, we adopt benchmark values of the oscillation parameters consistent with current global-fit results~\cite{Esteban:2024eli}, with the best-fit values corresponding to the Normal Ordering (NO) scenario given in Table~\ref{tab:nufit_bestfit_1sigma}. The diagonal elements $\mathcal{F}_{\theta_{23}\theta_{23}}$, $\mathcal{F}_{\delta_{\mathrm{CP}}\delta_{\mathrm{CP}}}$, and $\mathcal{F}_{\Delta m^2_{31}\Delta m^2_{31}}$ exhibit characteristic structures with enhanced sensitivity around $L/E\approx500$ and $1500$~km/GeV. Among them, $\mathcal{F}_{\Delta m^2_{31}\Delta m^2_{31}}$ dominates by several orders of magnitude, reaching values of $\sim10^{6}$--$10^{7}$, indicating a substantially stronger sensitivity of the neutrino quantum state to variations in $\Delta m^2_{31}$. The enhanced sensitivity near these regions originates from the oscillatory phase structure of neutrino flavor evolution, where variations in the oscillation parameters lead to appreciable changes in the quantum state. The comparatively large sensitivity to $\Delta m^2_{31}$ is associated with its direct contribution to the oscillation phase governing neutrino propagation. In contrast, $\mathcal{F}_{\delta_{\mathrm{CP}}\delta_{\mathrm{CP}}}\sim\mathcal{O}(0.1)$ remains comparatively small over the considered range, reflecting the weaker sensitivity to the CP phase in the present setup.

The corresponding QCRBs exhibit the expected inverse behavior with respect to the QFI. The uncertainty associated with $\Delta m^2_{31}$ lies in the range $10^{-7}$--$10^{-5}$, while the uncertainty in $\theta_{23}$ is comparatively larger, varying approximately between $10^{-1}$ and $10^{2}$. The uncertainty associated with $\delta_{\mathrm{CP}}$ can reach values above $\mathcal{O}(10^{3})$ in certain regions. This hierarchy follows directly from the relative sensitivities encoded in the diagonal QFIM elements. These bounds represent the quantum-limited precision for individual-parameter estimation; the impact of parameter correlations on simultaneous estimation is considered separately below.

To establish the connection between the quantum-limited sensitivity and the information accessible through measured flavor probabilities, Fig.~\ref{fig:2} compares the standard deviations obtained from the diagonal elements of the QFI and CFI. While the QFI provides a measurement-independent benchmark for the maximum information that can, in principle, be extracted from the underlying quantum state, the CFI quantifies the information accessible through the flavor-probability measurement considered here. The comparison reveals a strong dependence of the quantum--classical gap on both the oscillation parameter and the baseline-to-energy ratio. For $\theta_{23}$, the QFI- and CFI-based uncertainties become relatively close around $L/E\approx500$ and $1500$~km/GeV. For $\delta_{\mathrm{CP}}$, the difference is substantially larger over significant portions of the considered range, although the two uncertainties become equal near $L/E\approx1069$~km/GeV. In contrast, the QFI- and CFI-based uncertainties for $\Delta m^2_{31}$ remain comparatively close over most of the considered range, with more noticeable deviations around $L/E\approx500$ and $1500$~km/GeV.

For example, at $L/E=500$~km/GeV, the QFI- and CFI-based uncertainties are approximately $0.255$ and $0.260$ for $\theta_{23}$, $2.54$ and $19.7$ for $\delta_{\mathrm{CP}}$, and $7.9\times10^{-4}$ and $3.9\times10^{-3}$ for $\Delta m^2_{31}$, respectively. These results demonstrate that the extent to which the experimentally accessible probability information approaches the quantum limit is strongly parameter dependent. In particular, a smaller QFI--CFI gap indicates that the considered flavor-probability measurement extracts a larger fraction of the quantum information available for the corresponding parameter, whereas a larger gap indicates that a substantial amount of the quantum information is not captured by this measurement. Thus, the QFI should not be viewed as a substitute for the experimentally relevant CFI; rather, it provides a quantum-information benchmark against which the information extracted by a specific measurement can be assessed. This complementary role of the QFI and CFI provides a direct connection between the intrinsic quantum sensitivity of neutrino oscillations and the information accessible through flavor-probability measurements.

While the diagonal elements characterize the sensitivity to individual parameters, neutrino oscillations inherently constitute a multiparameter system, where the interplay between different oscillation parameters is encoded in the off-diagonal elements of the QFIM and CFIM. We therefore next investigate these off-diagonal elements and the corresponding parameter correlations, which determine how simultaneous estimation of the oscillation parameters is affected by their mutual dependence.}

\section{Multiparameter structure and intrinsic parameter correlations}
\label{sec:multi}
\begin{figure*}[!htb]
    \centering
    \includegraphics[width=0.49\textwidth]{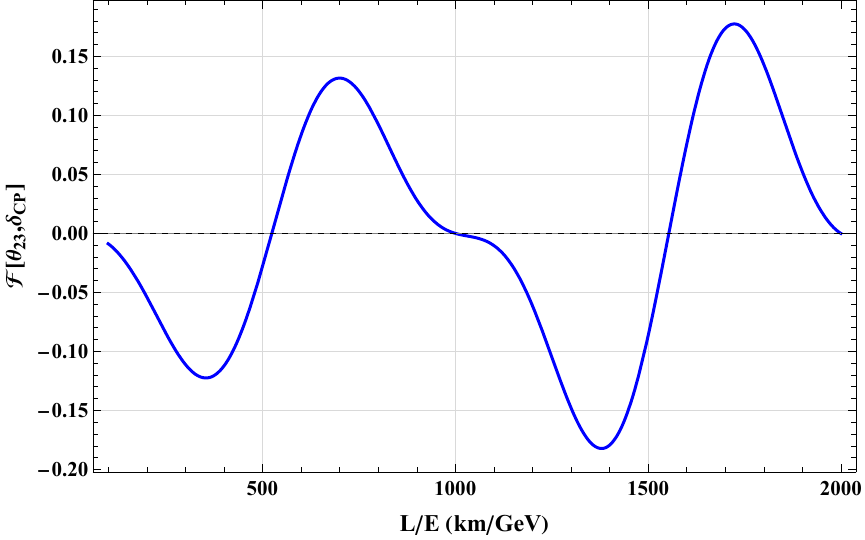}
    \includegraphics[width=0.49\textwidth]{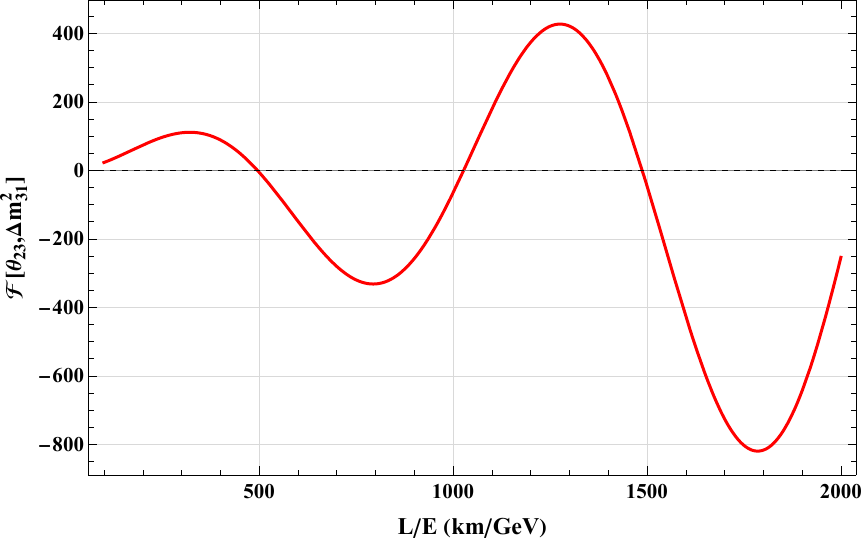}
    \includegraphics[width=0.49\textwidth]{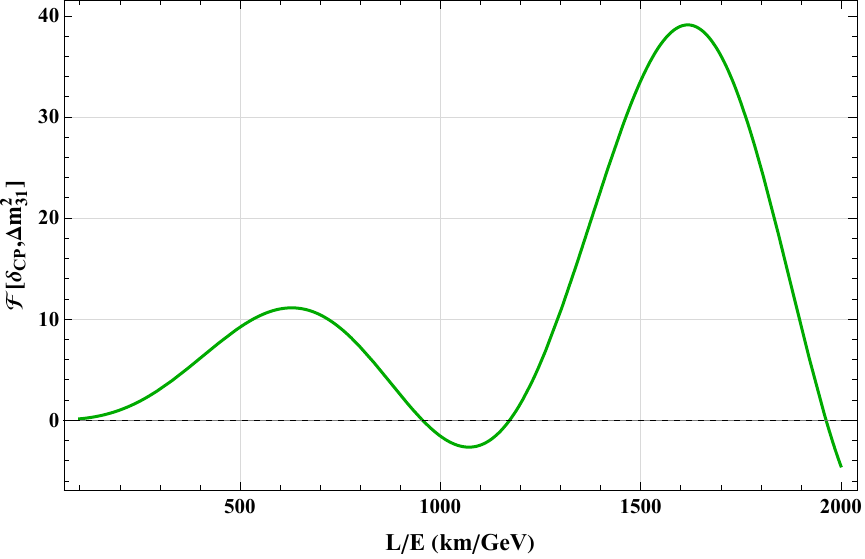}
     \includegraphics[width=0.49\textwidth]{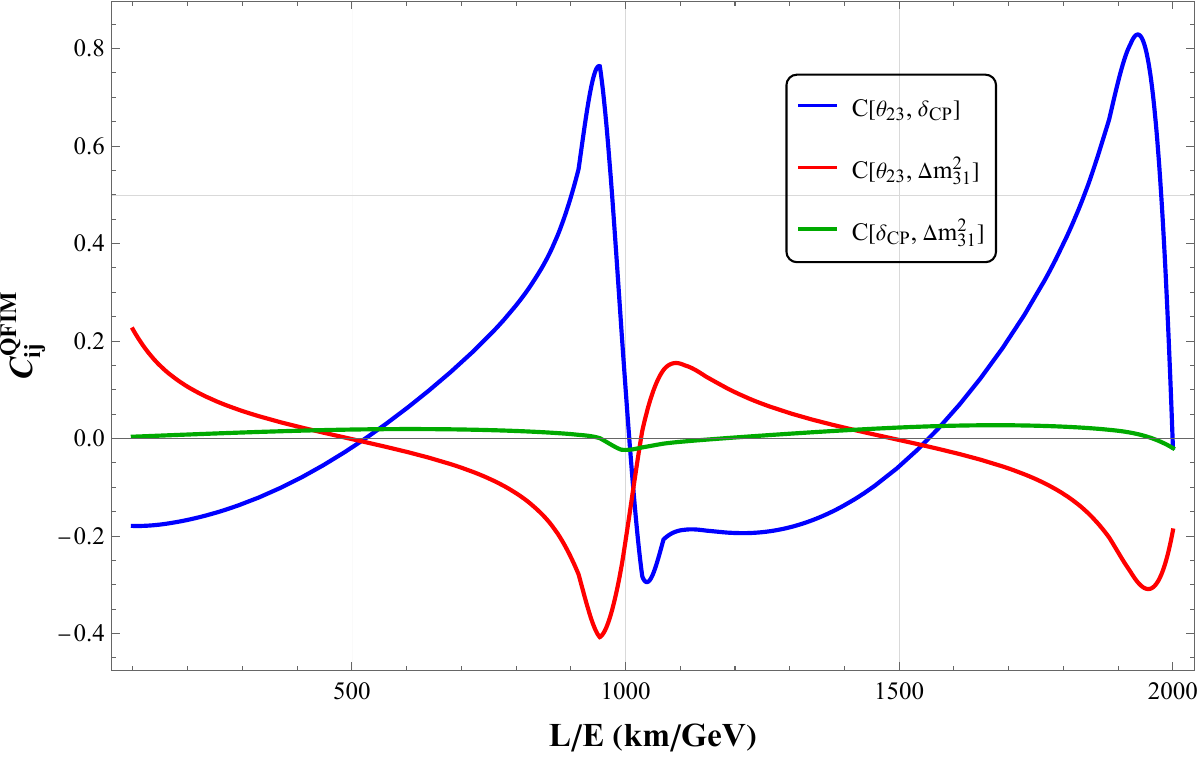}
    \caption{Quantum Fisher information of the Off-diagonal elements of the QFIM and the corresponding normalized parameter correlations (bottom-right panel) for different oscillation parameter pairs as functions of $L/E$. }
    \label{fig:3}
\end{figure*}

Fig.~\ref{fig:3} presents the off-diagonal elements of the QFIM and the corresponding normalized parameter correlations as functions of the baseline-to-energy ratio $(L/E)$. The off-diagonal elements $\mathcal{F}_{\theta_{23}\delta_{\mathrm{CP}}}$, $\mathcal{F}_{\theta_{23}\Delta m^2_{31}}$, and $\mathcal{F}_{\delta_{\mathrm{CP}}\Delta m^2_{31}}$ characterize the joint response of the neutrino quantum state to simultaneous variations of different oscillation parameters. Unlike the diagonal elements, the off-diagonal elements can take both positive and negative values, reflecting changes in the relative alignment of the parameter-dependent quantum-state variations across different regions of $L/E$.

The element $\mathcal{F}_{\theta_{23}\delta_{\mathrm{CP}}}$ exhibits an oscillatory behavior over the considered range, varying approximately between $-0.20$ and $+0.16$ with multiple sign reversals. The element $\mathcal{F}_{\theta_{23}\Delta m^2_{31}}$ has the largest absolute magnitude among the off-diagonal elements, spanning approximately $-800$ to $+400$, while $\mathcal{F}_{\delta_{\mathrm{CP}}\Delta m^2_{31}}$ exhibits an intermediate magnitude, reaching values of about $+39$ near $L/E\approx1600$~km/GeV. These pronounced baseline-dependent variations demonstrate that the simultaneous sensitivity to the oscillation parameters cannot be fully characterized by the diagonal QFI elements alone.

\begin{figure*}[!htb]
    \centering
    \includegraphics[width=0.32\linewidth,height=4cm]{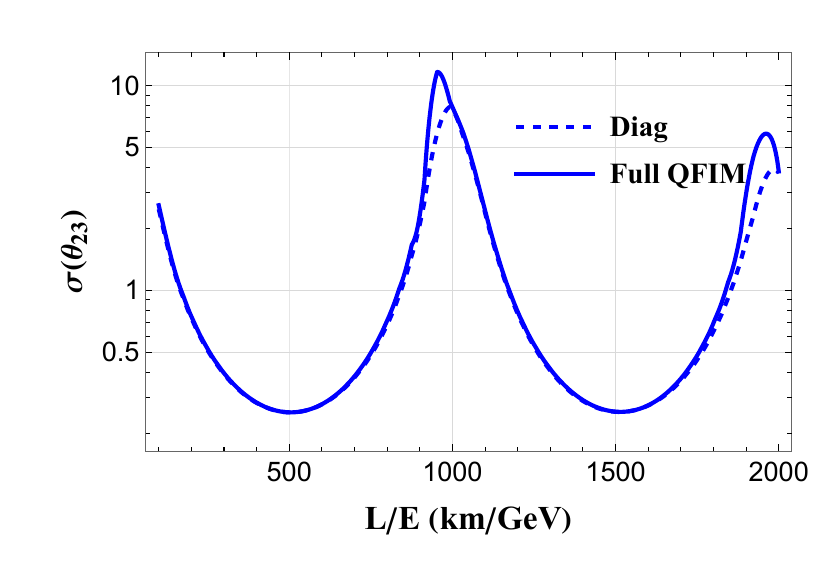}
    \includegraphics[width=0.32\linewidth,height=4cm]{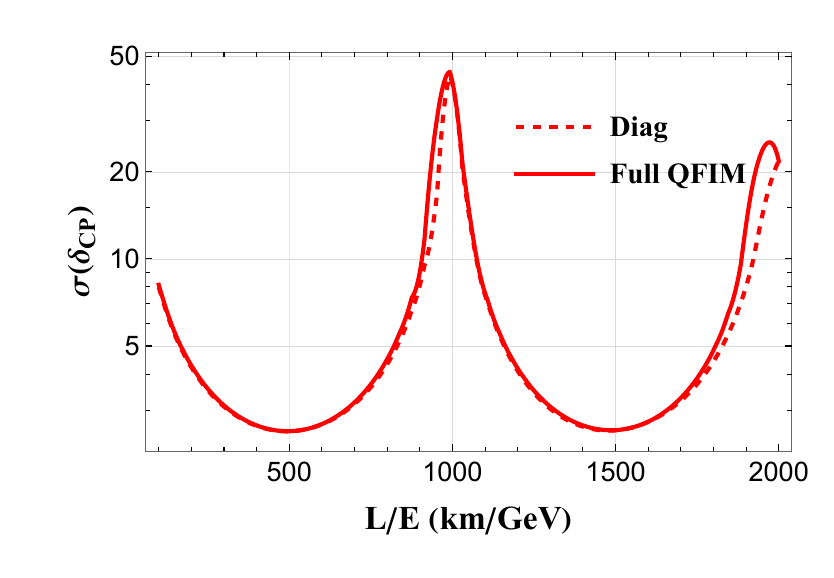}
    \includegraphics[width=0.32\linewidth,height=4cm]{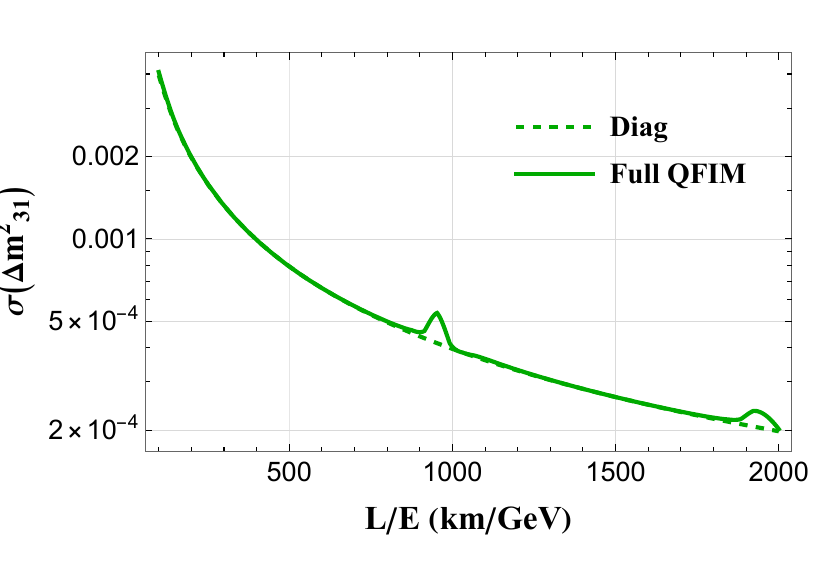}
    \caption{
Comparison of the standard deviations obtained using only the diagonal QFI elements (dashed curves) and the full QFIM including off-diagonal contributions (solid curves) for the oscillation parameters $\theta_{23}$, $\delta_{\mathrm{CP}}$, and $\Delta m^2_{31}$ as functions of $L/E$.
}
    \label{fig:4}
\end{figure*}

\begin{figure*}[!htb]
    \centering
    \includegraphics[width=0.49\textwidth]{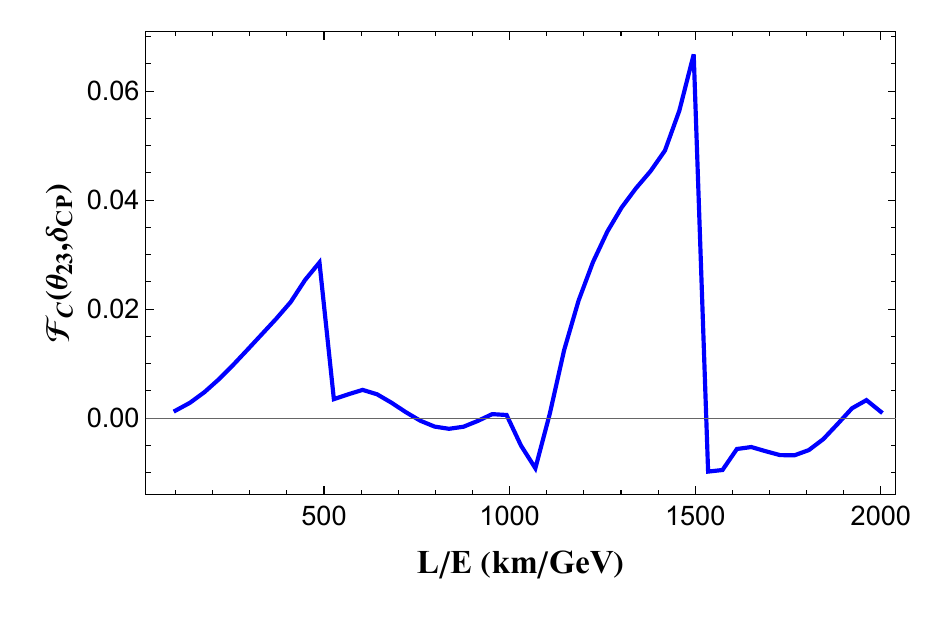}
    \includegraphics[width=0.49\textwidth]{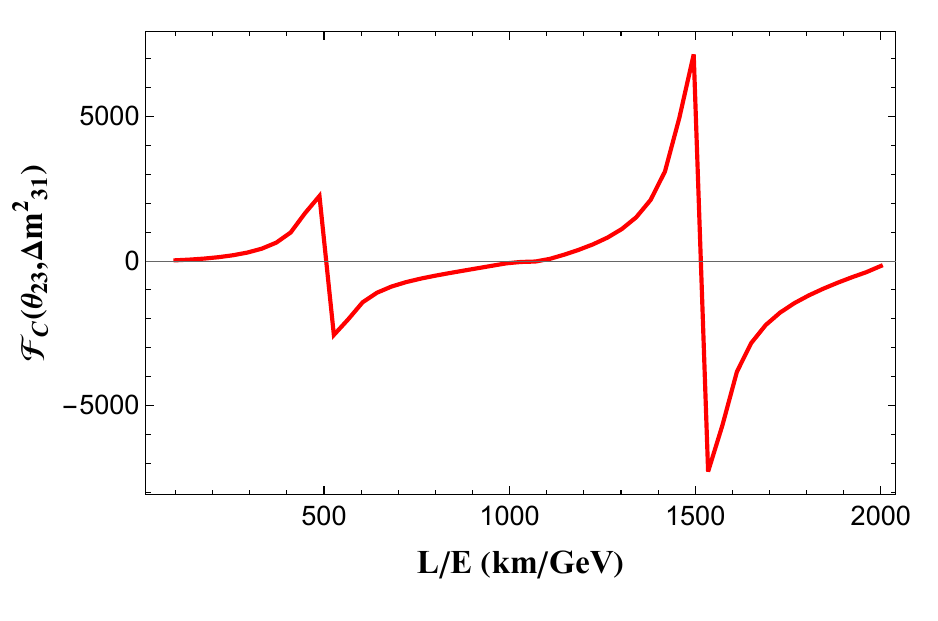}
    \includegraphics[width=0.49\textwidth]{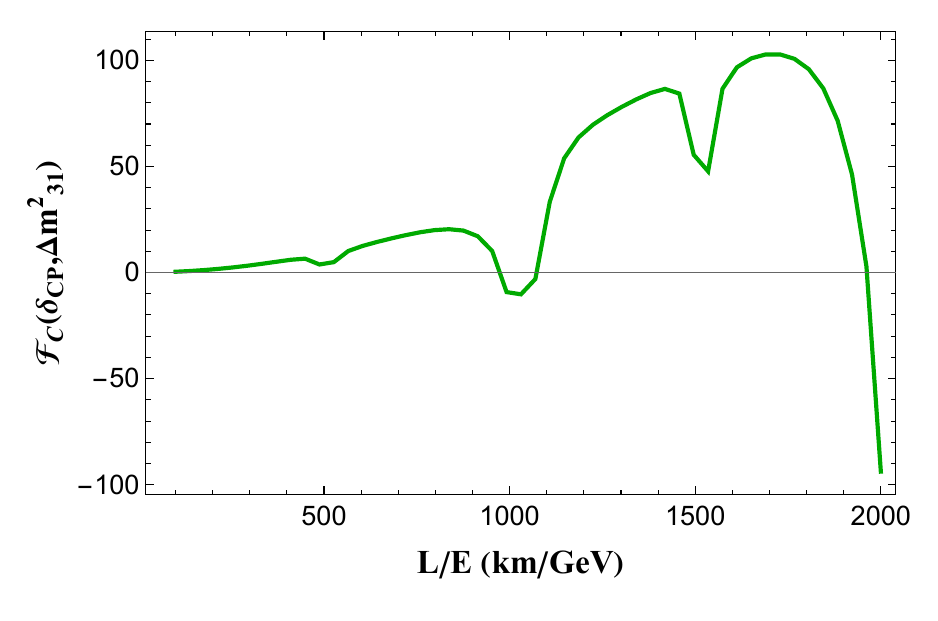}
     \includegraphics[width=0.49\textwidth]{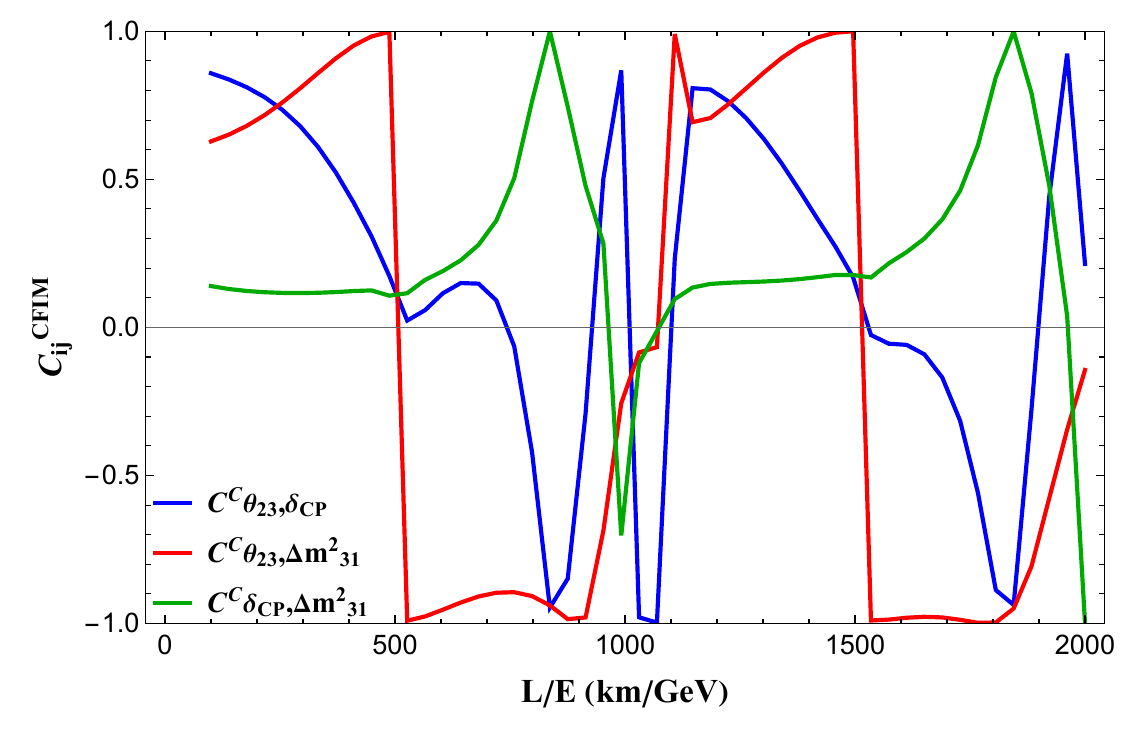}
    \caption{Classical Fisher information of the off-diagonal elements of the CFIM and the corresponding normalized parameter correlations (bottom-right panel) for different oscillation parameter pairs as functions of $L/E$. }
    \label{fig:5}
\end{figure*}

However, the magnitude of the raw QFI of the off-diagonal QFIM elements alone does not directly quantify the intrinsic correlations between oscillation parameters. To isolate the intrinsic correlation structure, we therefore consider the normalized coefficients
\begin{equation}
C_{ij}=\frac{\mathcal{F}_{ij}}
{\sqrt{\mathcal{F}_{ii}\mathcal{F}_{jj}}},
\end{equation}
shown in the bottom-right panel of Fig.~\ref{fig:3}. The sign of the correlation coefficient $C_{ij}$ reflects the nature of the relationship between the corresponding oscillation parameters. Positive values correspond to correlated variations of the parameters, while negative values indicate anti-correlated variations. The magnitude $|C_{ij}|$ quantifies the strength of the correlation, with larger values indicating stronger parameter interdependence. The $(\theta_{23},\delta_{\mathrm{CP}})$ pair exhibits pronounced intrinsic correlations, reaching values close to unity in specific $L/E$ regions, indicating that variations in these parameters can produce highly aligned modifications of the neutrino quantum state. In contrast, the correlation between $(\delta_{\mathrm{CP}},\Delta m^2_{31})$ remains comparatively weak over most of the considered range, implying a relatively weaker local interdependence between these parameters, although localized regions of enhanced correlation are present. The pair $(\theta_{23},\Delta m^2_{31})$ exhibits moderate-to-strong correlations with noticeable sign changes, reflecting transitions between aligned and oppositely aligned parameter-induced variations of the quantum state.

The present analysis may also have implications for future long-baseline accelerator experiments, which rely primarily on the $\nu_\mu\rightarrow\nu_e$ appearance channel. Since such experiments probe specific regions of the $L/E$ parameter space determined by their baselines and neutrino energy spectra, the correlation structure presented here can provide useful information regarding the simultaneous determination of oscillation parameters. In particular, regions where the correlation coefficients are small correspond to weaker local parameter interdependence and may therefore be associated with reduced parameter correlations in simultaneous estimation of $\theta_{23}$, $\delta_{\mathrm{CP}}$, and $\Delta m^2_{31}$. Consequently, the QFIM framework can serve as a complementary tool for identifying regions of enhanced sensitivity and characterizing the correlation structure among oscillation parameters in long-baseline oscillation studies.

As an illustrative experimental reference, we consider the DUNE experiment, which has a baseline of approximately $1300$~km and a neutrino flux concentrated at energies of a few GeV, corresponding to an $L/E$ region of several hundred km/GeV. For $E\simeq2.5$~GeV, for example, $L/E\simeq520$~km/GeV. From Fig.~\ref{fig:3} (bottom-right panel), the QFIM correlations in this region are relatively small compared with their larger values at other $L/E$ values, indicating comparatively weak local parameter interdependence among $\theta_{23}$, $\delta_{\mathrm{CP}}$, and $\Delta m^2_{31}$. This illustrates how the QFIM framework can provide a complementary quantum-information perspective on parameter correlations in an $L/E$ region relevant to long-baseline accelerator experiments. A quantitative assessment of the DUNE sensitivity, however, would require the full energy-dependent event spectrum together with flux, cross-section, detector-response, background, and systematic effects, which are beyond the scope of the present analysis.

Since the complete parameter uncertainties are governed by the inverse of the full QFIM,
\begin{equation}
    \mathrm{Cov}(\theta_\alpha,\theta_\beta)
    \geq
    \left(\mathcal{F}^{-1}\right)_{\alpha\beta},
\end{equation}
the off-diagonal structure of $\mathcal{F}$ can modify the marginal variances beyond the diagonal-only estimate $1/\mathcal{F}_{\alpha\alpha}$. To quantify the impact of parameter correlations on the achievable precision, we compare the standard deviations obtained from the full QFIM,
\begin{equation}
    \sigma(\theta_\alpha)
    =
    \sqrt{\left(\mathcal{F}^{-1}\right)_{\alpha\alpha}},
\end{equation}
with those obtained from the diagonal elements alone as functions of $L/E$ (see Fig.~\ref{fig:4}).

For both $\sigma(\theta_{23})$ and $\sigma(\delta_{\mathrm{CP}})$, the diagonal-only and full-QFIM estimates exhibit minima around $L/E\approx500$ and $1500$~km/GeV, consistent with the regions where the corresponding diagonal QFI elements attain their maxima. The two estimates remain close over most of the considered range, although noticeable deviations appear around $L/E\approx1000$~km/GeV and for $L/E\gtrsim1700$~km/GeV. These deviations indicate that the correlations among the oscillation parameters identified in Fig.~\ref{fig:3} can modify the marginal precision in specific regions, even when their effect remains moderate over most of the parameter space.

In contrast, $\sigma(\Delta m^2_{31})$ decreases monotonically over the considered $L/E$ interval, while the diagonal-only and full-QFIM estimates remain nearly indistinguishable throughout the range. This indicates that, for the benchmark considered here, the off-diagonal QFIM elements have only a minor effect on the marginal uncertainty associated with $\Delta m^2_{31}$. Thus, although Fig.~\ref{fig:3} reveals nonzero correlations involving $\Delta m^2_{31}$, these correlations do not lead to an appreciable modification of its marginal precision in the present analysis.

The QFIM characterizes the information geometry of the underlying neutrino quantum state, whereas the classical Fisher information depends on the measurement used to extract information from that state. To examine the parameter correlations accessible through flavor-probability measurements, Fig.~\ref{fig:5} presents the off-diagonal elements of the CFIM and the corresponding normalized CFI correlations as functions of $L/E$. The off-diagonal CFIM elements exhibit pronounced $L/E$ dependence and several sign changes, indicating that the probability distributions retain a nontrivial multiparameter correlation structure. In particular, $\mathcal{F}^{C}_{\theta_{23}\Delta m^2_{31}}$ shows the largest variation in magnitude, while $\mathcal{F}^{C}_{\theta_{23}\delta_{\mathrm{CP}}}$ and $\mathcal{F}^{C}_{\delta_{\mathrm{CP}}\Delta m^2_{31}}$ exhibit comparatively smaller but strongly $L/E$-dependent structures.

The normalized CFI correlations in the bottom-right panel further demonstrate that the strength and sign of the accessible parameter correlations vary substantially across the considered $L/E$ range. The correlations can approach $|C^C_{ij}|\simeq1$ in localized regions, indicating strong interdependence between the corresponding parameter responses in the flavor-probability measurement, while they become much smaller in other regions. The rapid changes and sign reversals observed in the CFI correlations reflect changes in the relative sensitivity of the measured probabilities to the oscillation parameters. Since the CFI is measurement dependent, these correlation patterns need not coincide with those obtained from the QFIM in Fig.~\ref{fig:3}. The comparison therefore distinguishes the intrinsic quantum correlation structure of the neutrino state from the parameter correlations that are accessible through the chosen flavor-probability measurement.

Taken together, Figs.~\ref{fig:3} and~\ref{fig:5} show that parameter correlations are present both at the quantum-state level and at the level of experimentally accessible flavor probabilities, but their detailed $L/E$ dependence can differ. The QFIM thus provides the measurement-independent quantum benchmark, while the CFIM reveals how the chosen probability measurement maps this quantum information into an experimentally accessible parameter-correlation structure.

\section{Information Geometry of Probability-Degenerate Solutions}

\label{sec:octant}

\begin{figure}[!htb]
    \centering
    \includegraphics[width=0.42
    \textwidth, height=10cm]{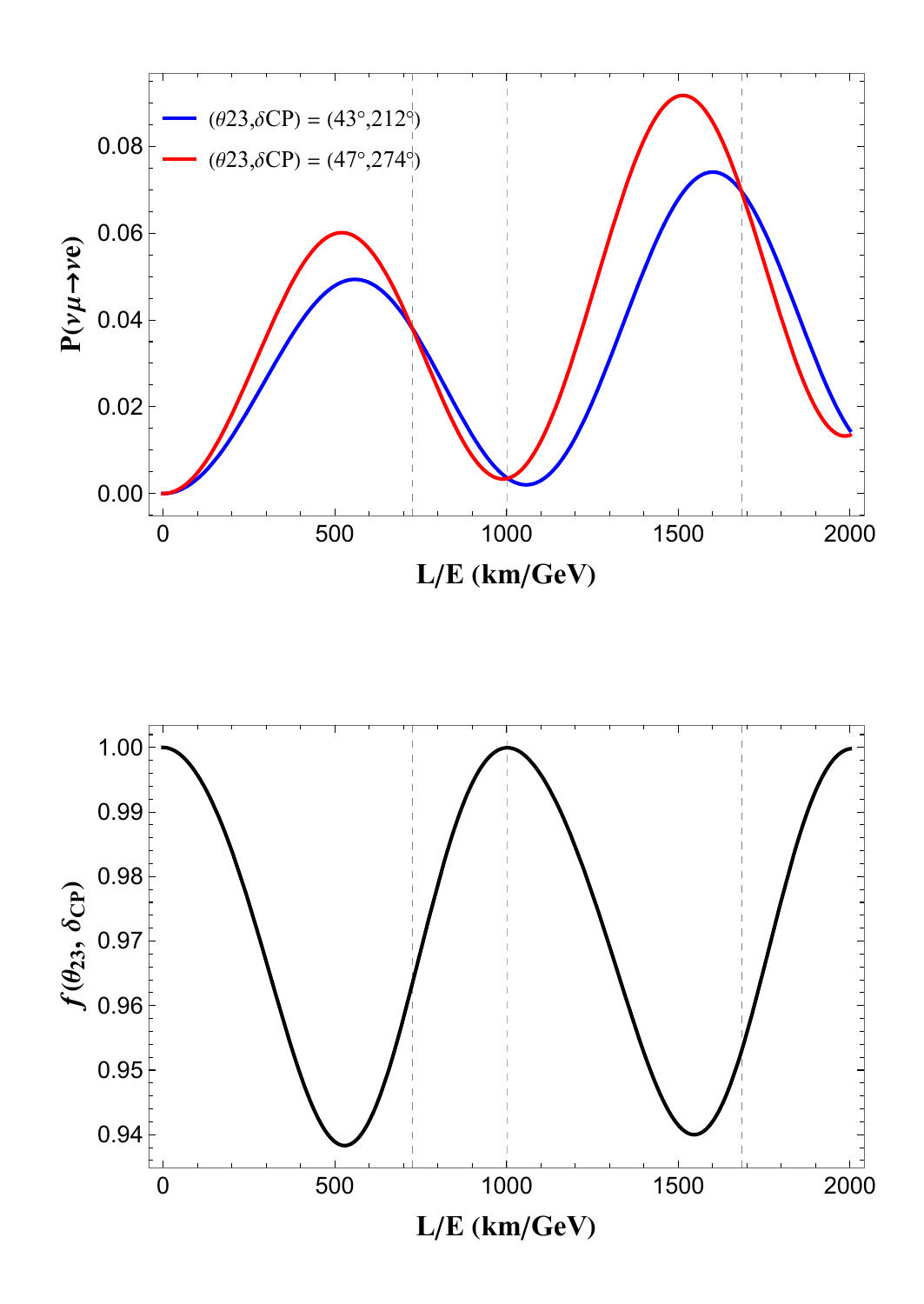}
  \caption{Upper panel: Appearance probability $P(\nu_\mu \rightarrow \nu_e)$ as a function of the baseline $L$ for two different oscillation parameter sets at $E_\nu=1~\mathrm{GeV}$. Lower panel: Corresponding quantum fidelity between the neutrino flavor states as a function of $L$. The vertical dashed lines denote representative degenerate baseline values where the degeneracy exists in appearance probability.}
    \label{fig:fidelity}
\end{figure}

In neutrino oscillation experiments, parameter degeneracies arise when different sets of oscillation parameters produce nearly identical transition probabilities. A well-known example is the eightfold degeneracy involving the atmospheric mixing angle, the CP-violating phase, and the neutrino mass ordering \cite{Barger:2001yr,Kajita:2006bt,Minakata:2010zn}. Since oscillation probabilities are classical observables constructed from quantum amplitudes, different parameter sets can produce nearly indistinguishable experimental signatures while corresponding to different underlying quantum states.

\begin{figure*}[!htb]
\centering
\includegraphics[width=0.32\textwidth]{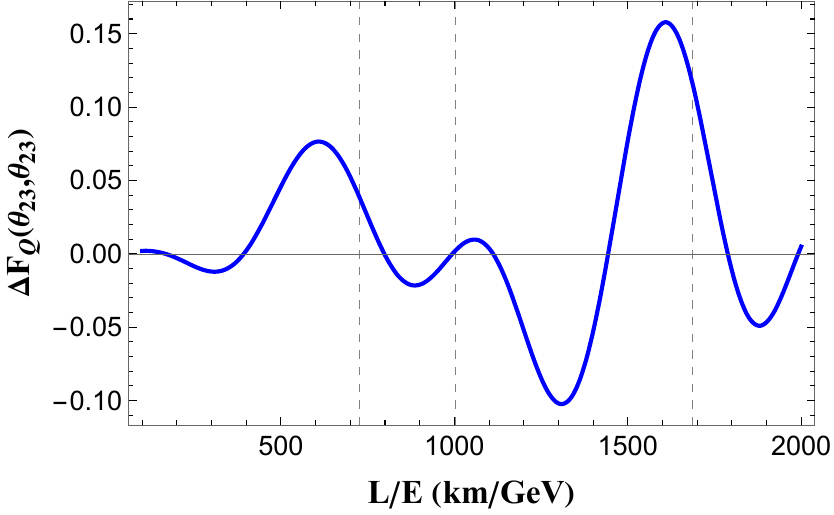}
\includegraphics[width=0.32\textwidth]{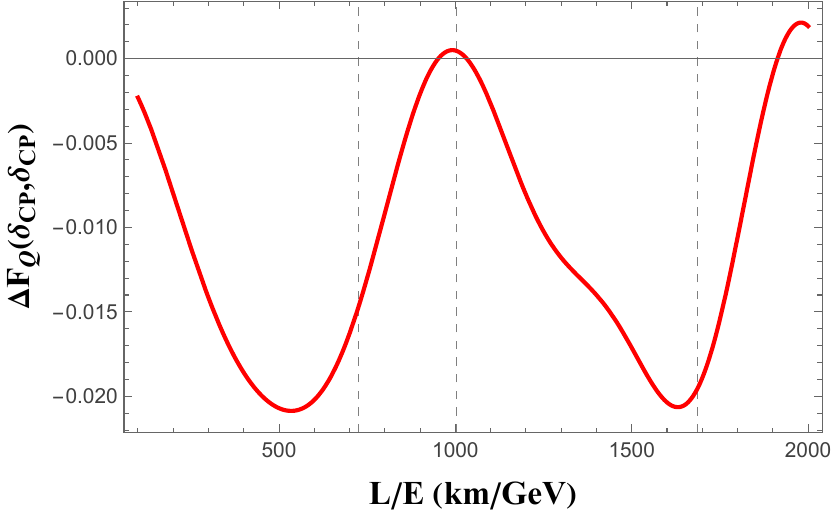}
\includegraphics[width=0.32\textwidth]{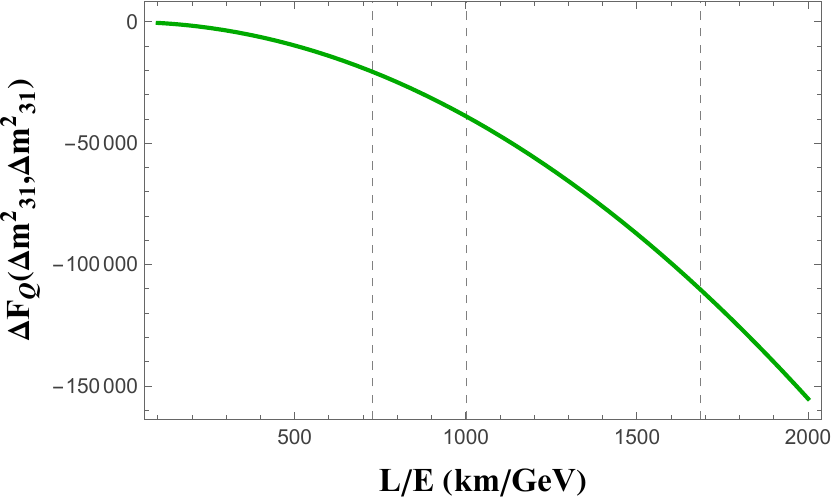}
\includegraphics[width=0.32\textwidth]{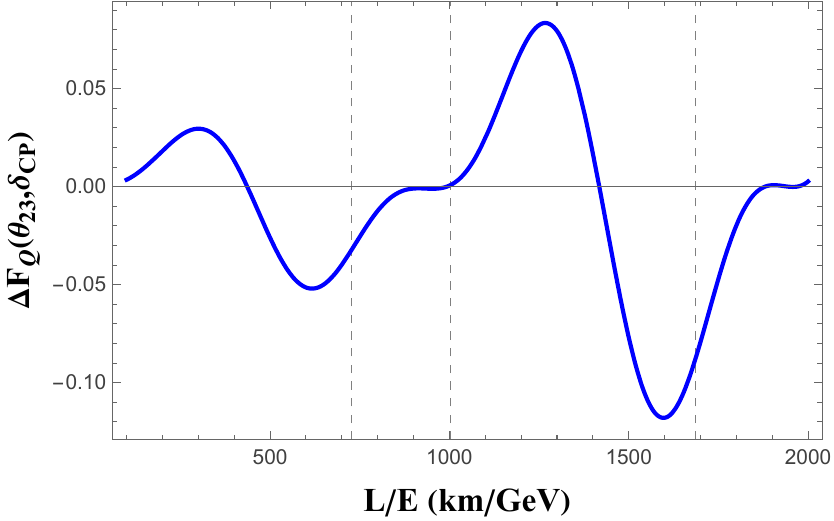}
\includegraphics[width=0.32\textwidth]{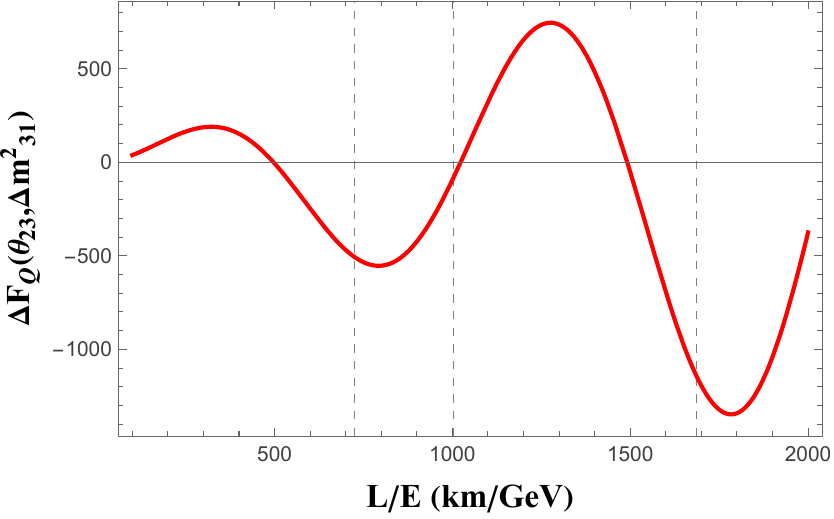}
\includegraphics[width=0.32\textwidth]{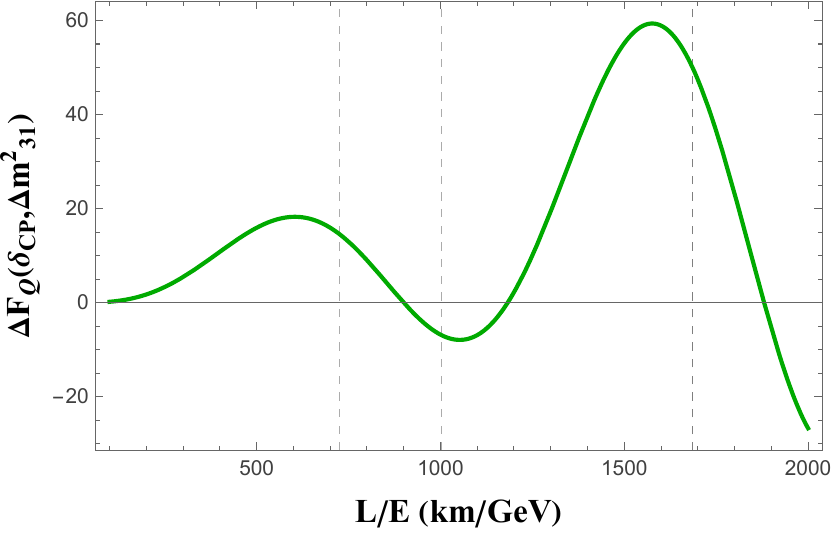}

\caption{Differences in the diagonal (top row) and off-diagonal (bottom row) elements of the quantum Fisher information matrix between two parameter sets, $(\theta_{23},\delta_{\rm CP})=(43^\circ,212^\circ)$ and $(47^\circ,274^\circ)$, that are degenerate in the $\nu_\mu\rightarrow\nu_e$ appearance probability, as a function of $L/E$. The vertical dashed lines indicate the corresponding $P_{\mu e}$-degenerate points at $L/E\simeq725$, $1002$, and $1685$~km/GeV.}
\label{fig:deltaqfi}
\end{figure*}

To investigate whether probability-degenerate solutions also possess identical quantum-information properties, we consider two parameter sets $\boldsymbol{\theta}_{1}$ and $\boldsymbol{\theta}_{2}$ satisfying
\begin{equation}
P_{\mu e}(\boldsymbol{\theta}_{1})
\approx
P_{\mu e}(\boldsymbol{\theta}_{2}).
\end{equation}
As an illustrative example, we consider
\begin{align}
(\theta_{23},\delta_{\mathrm{CP}})
&=
(43^\circ,212^\circ),
\qquad
(47^\circ,274^\circ),
\end{align}
with $\Delta m^2_{31}$ fixed. At $E=1~\mathrm{GeV}$, the two parameter sets yield identical $\nu_\mu\rightarrow\nu_e$ appearance probabilities at three representative baseline values, $L\simeq725$, $1002$, and $1685$~km, as shown in Fig.~\ref{fig:fidelity}. More precisely, these probability-degenerate points correspond to values of $L/E$ at which
\begin{equation}
P_{\mu e}^{(1)}=P_{\mu e}^{(2)}
\end{equation}
up to numerical precision ($\sim10^{-5}$). The quantum distinguishability of the corresponding flavor states is quantified by the fidelity \cite{Jozsa:1994qja},
\begin{equation}
f(1,2)
=
\left|
\langle
\psi(\theta_{23}^{(1)},\delta_{\mathrm{CP}}^{(1)})
|
\psi(\theta_{23}^{(2)},\delta_{\mathrm{CP}}^{(2)})
\rangle
\right|^{2}.
\end{equation}

\begin{figure*}[!htb]
\centering
\centering
\includegraphics[width=0.32\textwidth]{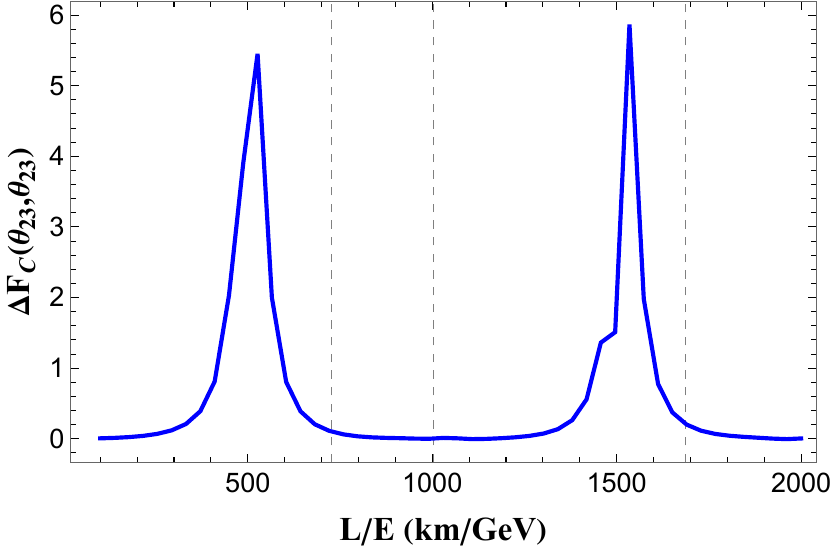}
\includegraphics[width=0.32\textwidth]{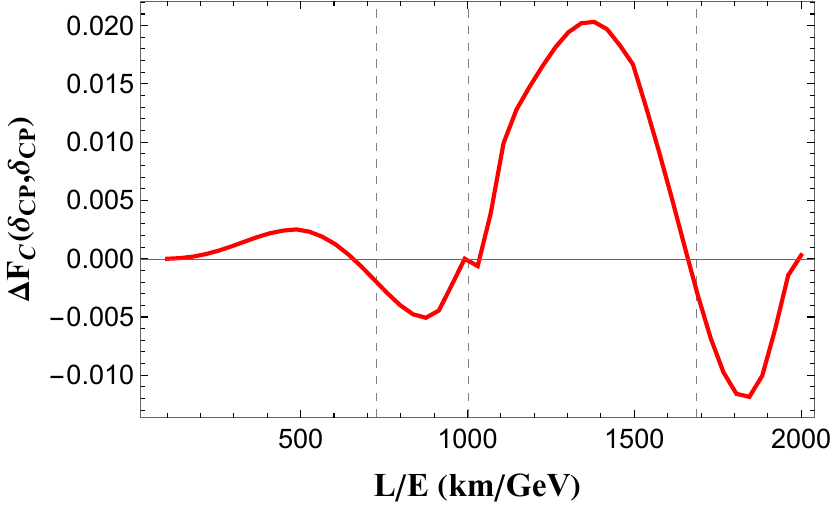}
\includegraphics[width=0.32\textwidth]{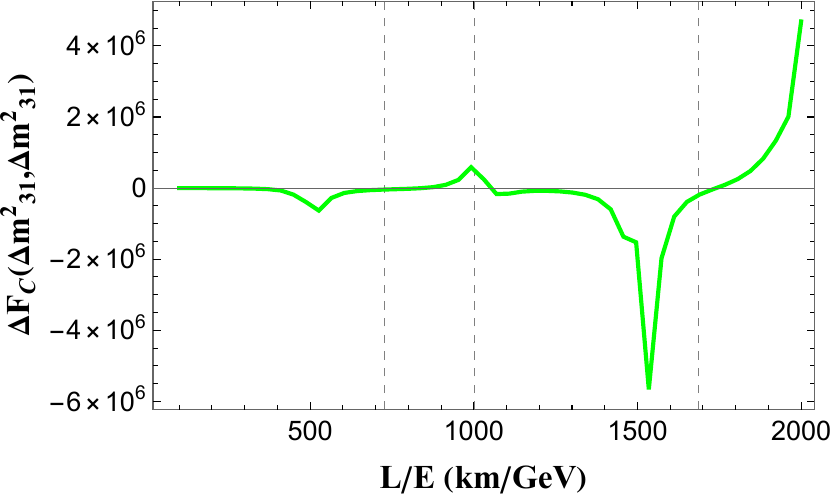}
\includegraphics[width=0.32\textwidth]{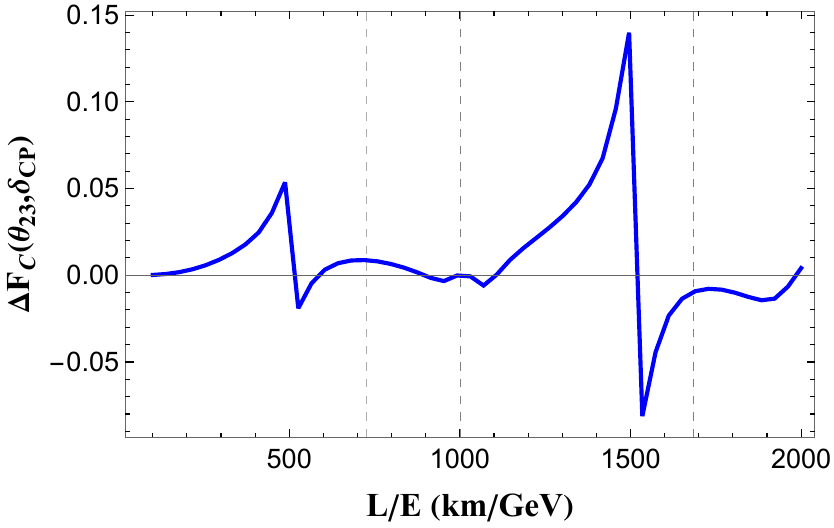}
\includegraphics[width=0.32\textwidth]{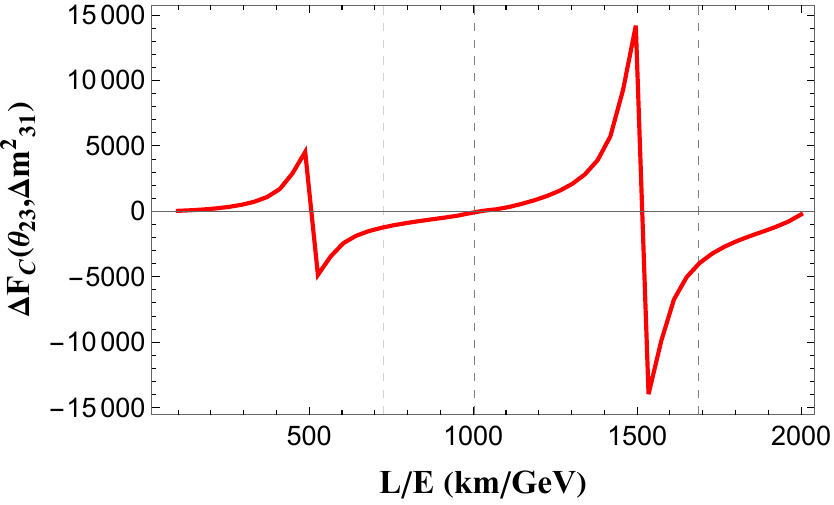}
\includegraphics[width=0.32\textwidth]{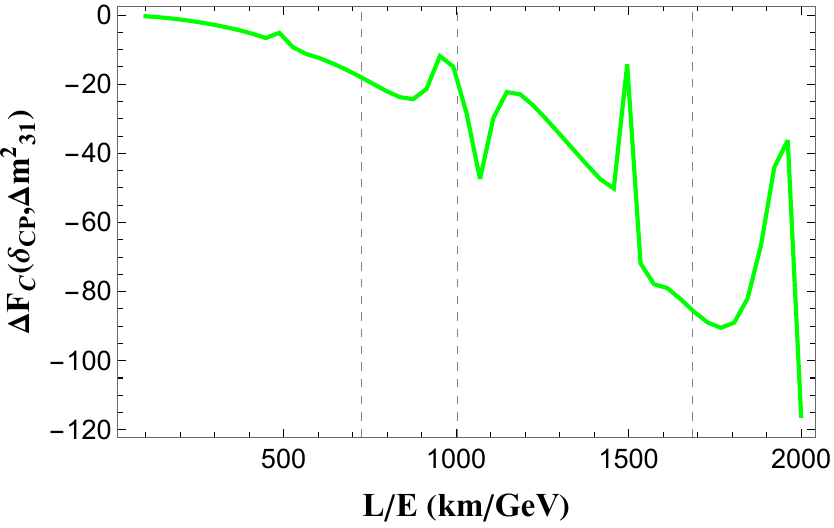}
\caption{Differences in the diagonal (top row) and off-diagonal (bottom row) elements of the classical Fisher information matrix between two parameter sets, $(\theta_{23},\delta_{\rm CP})=(43^\circ,212^\circ)$ and $(47^\circ,274^\circ)$, that are degenerate in the $\nu_\mu\rightarrow\nu_e$ appearance probability, as a function of $L/E$. The vertical dashed lines indicate the corresponding $P_{\mu e}$-degenerate points at $L/E\simeq725$, $1002$, and $1685$~km/GeV.}
\label{fig:deltacfi}
\end{figure*}

At the three $P_{\mu e}$-degenerate points, we obtain a fidelity of approximately $f\simeq0.953$, indicating a large overlap between the two quantum states while still allowing them to remain distinguishable. Thus, equality of the appearance probability does not imply equality of the underlying quantum states.

Although the fidelity establishes the existence of quantum distinguishability, it is a global overlap measure and does not directly reveal which parameter directions carry different local sensitivity. The QFIM provides complementary information by characterizing the local information geometry of the quantum state with respect to the oscillation parameters. To this end, we define the element-wise QFIM difference
\begin{equation}
\Delta\mathcal{F}_{ij}(L/E)
\equiv
\mathcal{F}_{ij}^{(1)}(L/E)
-
\mathcal{F}_{ij}^{(2)}(L/E),
\label{eq:deltaF}
\end{equation}
where $\mathcal{F}_{ij}^{(k)}$ denotes the QFIM evaluated at $\boldsymbol{\theta}_k$. A nonzero value of $\Delta\mathcal{F}_{ij}$ at a $P_{\mu e}$-degenerate point indicates that the two parameter sets possess different local quantum-information geometries, despite producing the same appearance probability at that point. Figure~\ref{fig:deltaqfi} displays the six independent elements of $\Delta\mathcal{F}_{ij}$ as functions of $L/E$, with the vertical lines indicating the representative $P_{\mu e}$-degenerate points.

Examining the diagonal elements (top row of Fig.~\ref{fig:deltaqfi}), we find that all three exhibit nonzero differences at the representative $P_{\mu e}$-degenerate baselines. The $\theta_{23}$ element (blue) exhibits an oscillatory behavior, varying approximately between $-0.11$ and $+0.16$, indicating that the two parameter sets differ in their local sensitivity to the atmospheric mixing angle. The $\delta_{\mathrm{CP}}$ element is smaller in magnitude, varying approximately between $-0.021$ and $+0.002$, but remains nonzero at the degenerate points. The $\Delta m^2_{31}$ element is qualitatively different, showing a predominantly negative difference that grows in magnitude with $L/E$ and reaches approximately $-1.6\times10^{4}$ at $L/E=2000$~km/GeV. These results demonstrate that the two parameter sets can possess markedly different local sensitivities to the oscillation parameters even at values of $L/E$ where their $\nu_\mu\rightarrow\nu_e$ appearance probabilities coincide.

The off-diagonal elements (bottom row of Fig.~\ref{fig:deltaqfi}) encode differences in the local cross-parameter information structure. The $\theta_{23}$--$\delta_{\mathrm{CP}}$ element exhibits an oscillatory behavior, varying approximately between $-0.12$ and $+0.08$. The $\theta_{23}$--$\Delta m^2_{31}$ element shows a much larger variation, reaching approximately $+7\times10^{2}$ and decreasing to about $-1.2\times10^{3}$ over the considered range. The $\delta_{\mathrm{CP}}$--$\Delta m^2_{31}$ element varies approximately between $-30$ and $+60$. These pronounced variations demonstrate that the two probability-degenerate solutions also possess different local cross-parameter information structures. In particular, the nonzero differences in the off-diagonal elements at the representative $P_{\mu e}$-degenerate points show that the parameter correlations associated with the two solutions are not identical. Thus, the QFIM reveals that the two probability-degenerate solutions correspond to different local sensitivity and correlation structures in the full multiparameter space.

{\color{black} Since the QFIM characterizes the measurement-independent information geometry of the quantum state, it is also important to examine whether analogous differences are present in the information accessible through the measured flavor probabilities. We therefore define the corresponding element-wise CFIM difference,
\begin{equation}
\Delta\mathcal{F}^{C}_{ij}(L/E)
\equiv
\mathcal{F}^{C,(1)}_{ij}(L/E)
-
\mathcal{F}^{C,(2)}_{ij}(L/E),
\end{equation}
where $\mathcal{F}^{C,(k)}_{ij}$ denotes the CFIM evaluated at the two parameter sets. Figure~\ref{fig:deltacfi} shows the six independent elements of this difference. 

An interesting feature of Fig.~\ref{fig:deltacfi} is that the differences in the CFIM elements are strongly parameter and $L/E$ dependent. While some elements exhibit pronounced differences between the two probability-degenerate solutions, others remain close to zero over extended regions of $L/E$. This indicates that the two solutions can have nearly identical local classical information content along particular parameter directions, even though their overall information structures are different. In particular, a small value of $\Delta\mathcal{F}^{C}_{ij}$ does not imply that the two underlying quantum states are identical; rather, it indicates that the chosen flavor-probability measurement provides similar local information about the corresponding parameter combination for the two solutions.

The comparison with Fig.~\ref{fig:deltaqfi} is therefore particularly instructive. The QFIM differences quantify changes in the intrinsic local quantum-information geometry, whereas the CFIM differences quantify which of these changes are reflected in the information accessible through the flavor-probability measurement. Consequently, a large $\Delta\mathcal{F}_{ij}$ need not be accompanied by a comparably large $\Delta\mathcal{F}^{C}_{ij}$. This demonstrates that the distinction between probability-degenerate solutions depends not only on the underlying quantum states but also on the measurement through which the states are probed. Rather than lifting the probability degeneracy itself, the QFIM and CFIM therefore reveal complementary information about the distinct local parameter sensitivities and correlations associated with probability-degenerate solutions.
}

\section{Conclusion}
\label{sec:conclusion}

In this work, we have investigated the multiparameter quantum-information structure of three-flavor neutrino oscillations using the QFIM, considering $\theta_{23}$, $\delta_{\mathrm{CP}}$, and $\Delta m^2_{31}$ as the parameters of interest. We have analyzed the diagonal and off-diagonal QFIM elements, the corresponding quantum Cram\'er--Rao bounds, and the normalized parameter correlations as functions of the baseline-to-energy ratio $L/E$. We have further compared the quantum Fisher information with the CFI associated with the flavor-probability measurement.

The diagonal QFIM elements exhibit a clear parameter-dependent sensitivity structure. The information associated with $\Delta m^2_{31}$ is substantially larger than that associated with $\theta_{23}$ and $\delta_{\mathrm{CP}}$, while the corresponding quantum-limited uncertainties show the complementary behavior. For $\theta_{23}$ and $\delta_{\mathrm{CP}}$, enhanced sensitivity occurs around characteristic regions near $L/E\approx500$ and $1500$~km/GeV, whereas the sensitivity to $\Delta m^2_{31}$ exhibits a different, more continuously varying behavior. The comparison with the CFI shows that the experimentally accessible information depends strongly on both the oscillation parameter and $L/E$. In particular, the CFI uncertainty for $\theta_{23}$ can remain close to the QFI-based limit, whereas substantially larger differences are observed for $\delta_{\mathrm{CP}}$. The QFI therefore provides a measurement-independent quantum benchmark against which the information extracted by a specific flavor-probability measurement can be assessed.

The multiparameter analysis demonstrates that the diagonal information alone does not fully characterize neutrino parameter estimation. The off-diagonal QFIM and CFIM elements exhibit pronounced $L/E$ dependence and sign changes, reflecting nontrivial parameter correlations. The normalized correlations further show that a large raw off-diagonal information element does not necessarily correspond to a strong normalized correlation. The comparison between diagonal-only and full-QFIM uncertainties indicates that parameter correlations modify the marginal precision primarily in specific $L/E$ regions, while their effect remains relatively small over much of the considered range.

We have also investigated probability-degenerate solutions in the $(\theta_{23},\delta_{\mathrm{CP}})$ parameter space. The two parameter sets $(43^\circ,212^\circ)$ and $(47^\circ,274^\circ)$ produce identical $\nu_\mu\rightarrow\nu_e$ appearance probabilities at representative $L/E$ values around $725$, $1002$, and $1685$~km/GeV. Although the corresponding quantum states have a large overlap, with a fidelity of approximately $f\simeq0.953$, they are not identical. The QFIM differences between the two solutions reveal distinct local quantum-information geometries at the probability-degenerate points. The corresponding CFIM differences show a more parameter- and $L/E$-dependent behavior, with some elements exhibiting pronounced differences while others remain close to zero over extended regions. Thus, probability degeneracy does not imply identical quantum-state information geometry, nor does it imply identical local classical information accessible through a particular measurement.

Overall, the combined QFI--CFI analysis provides a complementary perspective on neutrino oscillation parameter estimation. The QFIM characterizes the intrinsic information content of the neutrino quantum state, while the CFI quantifies the information accessible through the specified flavor-probability measurement. These results motivate further studies incorporating realistic energy-dependent event spectra, detector effects, matter effects, and systematic uncertainties to assess the experimental implications of the quantum- and classical-information structures identified here.

\section*{Acknowledgements}
The authors gratefully acknowledge the support of the National Natural Science Foundation of China under Grant No. 12075059, as well as the start-up fund of USTC.

\bibliography{refer1}
\end{document}